\documentclass[aps, prl, reprint]{revtex4-1}

\usepackage{amsmath, amssymb, times, graphicx, mathrsfs, hyperref, array, bbm}
\usepackage[usenames,dvipsnames]{color}
\hypersetup{colorlinks=false}

\graphicspath{{./}{./figures/}}

\newcommand{\beq}{\begin{equation}}
\newcommand{\eeq}{\end{equation}}

\newcommand{\tr}{\text{tr}}

\newcommand{\intg}{\mathbb{Z}}

\newcommand{\id}{\mathbbm{1}}

\newcommand{\hlt}{\mathcal{H}}

\newcommand{\ve}{\varepsilon}

\newcommand{\vk}{\mathbf{k}}

\renewcommand{\vr}{\mathbf{r}}

\newcommand{\viz}{\emph{viz}}

\newcommand{\e}{\mathrm{e}}

\begin{document}
\title{Majorana Corner Modes in a Second-Order Kitaev Spin Liquid}

\author{Vatsal Dwivedi}
\email[E-mail: ]{vdwivedi@thp.uni-koeln.de}
\author{Ciar\'an Hickey}
\author{Tim Eschmann}
\author{Simon Trebst}
\affiliation{Institute for Theoretical Physics, University of Cologne, 50937 Cologne, Germany}

\date{\today}

\begin{abstract}
  Higher-order topological insulators are distinguished by the existence of topologically protected modes with codimension two or higher. 
  Here, we report the manifestation of a second-order topological insulator in a two dimensional frustrated quantum magnet, which exhibits topological corner modes. 
  Our exactly-solvable model is a generalization of the Kitaev honeycomb model to the Shastry-Sutherland lattice that, besides a chiral spin liquid phase, exhibits a gapped spin liquid with Majorana corner modes, which are protected by two mirror symmetries. 
  This second-order Kitaev spin liquid remains stable in the presence of thermal fluctuations and undergoes a finite-temperature phase transition evidenced in large-scale quantum Monte Carlo simulations.
\end{abstract}

\maketitle

The study of topological band theory for non-interacting electron systems has led to the advent of a plethora of topological insulators (TIs) \cite{kane-mele, fu-kane-mele,hasan-kane_TI,Qi2011,bhbook}. 
A central feature of these systems is the existence of gapless boundary modes, which are protected by the topology of the bulk bands, i.e. they cannot be gapped out by any deformation of the Hamiltonian which keeps the bulk gap open and preserves certain symmetries. 
These systems are termed `topological' because their low energy behavior is governed by a topological 
action, which is independent of microscopic details of the system such as the underlying lattice structure. 
However, this is strictly true only for \emph{strong TIs}, which are protected by time reversal and/or charge conjugation symmetries. 
In recent years, a class of more `fragile' variants of these phases, termed \emph{crystalline TIs}, has been explored.
For these systems, the boundary modes are protected only under Hamiltonian deformations that preserve certain lattice symmetries \cite{Fu2011tci}, and exist only on boundaries that are themselves invariant under these symmetries. 
Importantly, for these more fragile systems the crystal structure remains important even for the low-energy physics.

Recently, the family of crystalline TIs has been expanded by what are best called \emph{higher-order topological insulators} \cite{hughes-bernevig_quadrupole_insulators, hughes-bernevig_multipole_insulators,neupert_HOTI}. In this paradigm, an $n^\text{th}$-order TI is a $d$-dimensional insulator that exhibits topologically protected gapless modes only in $d-n$ spatial dimensions localized at the intersection of $n$ boundary planes, while the boundaries of codimension less than $n$ remain gapped. 
For instance, a second-order TI (SOTI) in two spatial dimensions is an insulator whose edge state itself is a one-dimensional TI, with zero modes localized only at the corners of the system.
Various crystalline symmetries have been invoked for the protection of the zero modes, including order-two lattice symmetries
\cite{hughes-bernevig_quadrupole_insulators, hughes-bernevig_multipole_insulators, brouwer_HOTI,brouwer_HOTI_classification} (such as mirror reflection, twofold rotation, or inversion symmetry)
or higher-order lattice symmetries such as $C_4$ rotation symmetry~\cite{song-fang_HOTI, neupert_HOTI}.
Inspired by this theoretical work, experimental realizations of higher-order TIs have been observed as phononic TI in a
cleverly designed mechanical metamaterial \cite{huber_HOTI_phonons} and as quantized quadrupolar TIs in electrical \cite{thomale_HOTI_electrical} and microwave \cite{bahl_HOTI_microwave} circuits, along with the recent discovery that elemental Bismuth is in fact a second-order TI \cite{neupert_HOTI_bismuth}.

In this manuscript, we introduce an exactly solvable microscopic spin model of a frustrated quantum magnet, which exhibits an analogue of the SOTI in a strongly interacting system. More precisely, our model exhibits spin liquid physics at low temperatures, with a fractionalization of its local degrees of freedom into itinerant Majorana fermions and a static $\intg_2$ gauge field. The band structure of the Majorana fermions reveals a phase diagram with not only a conventional Chern insulator, but also a 
SOTI with topologically protected corner modes. The former corresponds to the formation of a chiral spin liquid ground state, while the latter is the first instance of a {\em second-order} spin liquid. Both spin liquids describe states with spontaneously broken time-reversal symmetry, which are separated from the high-temperature paramagnet by a finite-temperature phase transition. We track this thermal phase transition and the prior spin fractionalization in various thermodynamic observables calculated via sign problem-free quantum Monte Carlo simulations of our spin model.

\begin{figure*}[t]
    \includegraphics[ width=\textwidth]{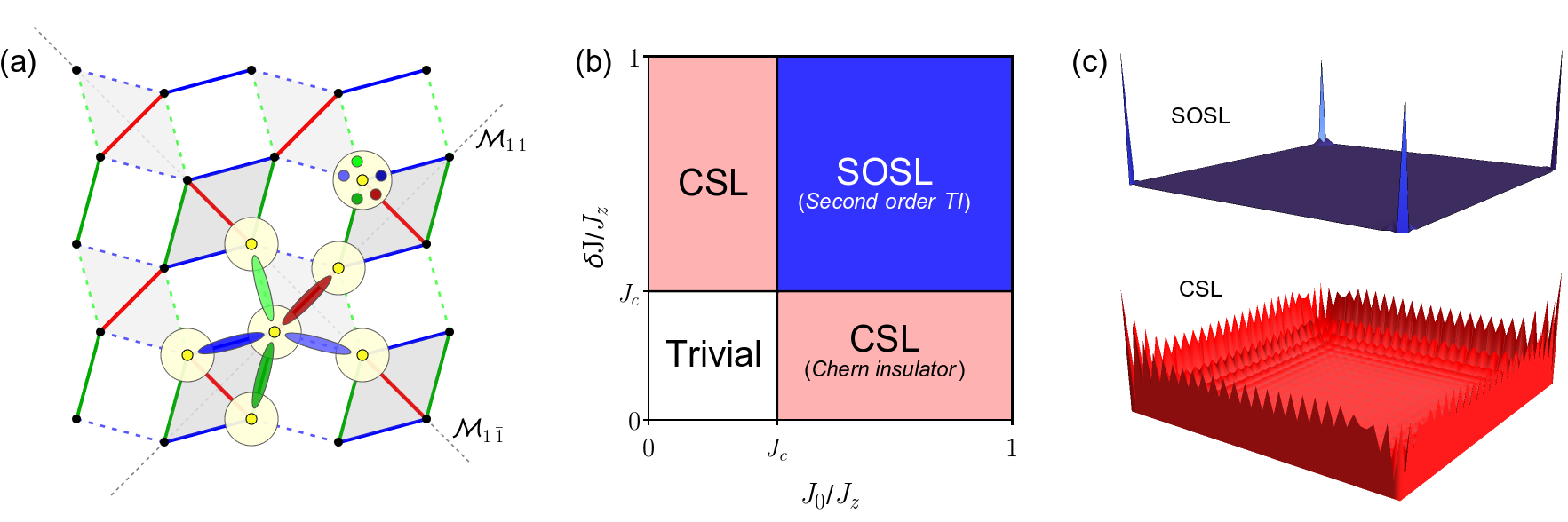}
    \caption{
      {\bf The higher-order Kitaev model on the Shastry-Sutherland lattice.}
      (a) The Shastry-Sutherland lattice, with the spin $x,y,z$ bonds for the 1D Kitaev chain depicted by solid blue, green and red lines, respectively, while the orbital $x,y$ bonds are depicted by dashed blue and green lines, respectively. The dark and light gray shading denotes the two kinds of plaquettes with couplings $J_0 + \delta J$ and $J_0 - \delta J$, respectively. The dotted gray lines denote the two mirror axes. 
      (b) The phase diagram as a function of average couplings and staggerings on the rhombi. Here, $J_c = \frac{1}{2\sqrt{2}}$. 
      (c) The wavefunctions for the four corner modes in the SOSL phase and an edge mode in the CSL phase. 
    }
    \label{fig:panel}
\end{figure*}

\paragraph{Microscopic Model.--}
We consider a higher-spin realization of the Kitaev honeycomb model \cite{Kitaev2006anyons} to the Shastry-Sutherland lattice \cite{shastry-sutherland} illustrated in Fig.~\ref{fig:panel}. This lattice is best known for the orthogonal-dimer model, which has been solved exactly by Shastry and Sutherland \cite{shastry-sutherland} and is a remarkably good description for the low-temperature physics of the transition metal oxide SrCu$_2$(BO$_3$)$_2$ \cite{Kageyama1999}. As a five-coordinated lattice, it shares an {\em odd} coordination number for every site with the tricoordinated honeycomb lattice, which is a crucial ingredient 
\footnote{Analytically tractable generalizations of the Kitaev model to lattices with an even coordination number have been put forward in Refs.~\onlinecite{yao_algebraic_spin_liquid,ryu_kitaev_diamond_lattice,ryu_kitaev_sqr_lattice}} 
to construct an exactly solvable Kitaev model. For the honeycomb Kitaev model, the tricoordination of the sites matches perfectly with the decomposition of the original spin-1/2 degrees of freedom into three ``bond Majorana fermions", which are recombined into $\intg_2$ gauge fields (assigned to the bonds), and one itinerant Majorana fermion. By analogy, a five-coordinated lattice asks for six Majorana fermions, which in principle span a Hilbert space of eight states. However, keeping in mind that the physical subspace of a Kitaev model needs a projection to precisely half of this Hilbert space, we are looking for constituent degrees of freedom that span a local Hilbert space of only four states. This can be achieved by either considering a $j=3/2$ spin degree of freedom or, alternatively, two coupled spin-1/2 degrees of freedom, such as spin and orbital degrees of freedom. Using the latter, we first define our microscopic model as 
\begin{align}
  \hlt = \sum_{\langle i,j \rangle} J_\gamma  (\tau_i^z \tau_j^z) \otimes (\sigma_i^\gamma \sigma_j^\gamma) + \sum_{( i,j )} J_\delta' \tau_i^\delta \tau_j^\delta \otimes \id, 
\end{align}
where the Pauli matrices $\sigma$ and $\tau$ denote the spin and orbital degrees of freedom and $\langle i,j \rangle$ and $( i,j )$ 
indicate couplings along the solid/dashed bonds in Fig.~\ref{fig:panel}a, respectively. Five different bond types that couple spin and orbital components $\gamma \in \{x,y,z\}$ and $\delta \in \{x,y\}$, respectively, are defined as marked in Fig.~\ref{fig:panel}a. 
We further allow a staggering of the couplings on the two kinds of rhombi (shaded in dark and light gray in Fig.~\ref{fig:panel}a).

In order to solve this model exactly, we first recast it into a Kitaev-like form by defining, for each site,  the $4\times 4$ anticommutating matrices 
\begin{align}
  & \Gamma^1 = \tau^x \otimes \id, \quad 
  \Gamma^2 = \tau^y \otimes \id, \quad 
  \Gamma^3 = \tau^z \otimes \sigma^x, \nonumber \\  
  & \qquad\qquad \Gamma^4 = \tau^z \otimes \sigma^y, \quad  
  \Gamma^5 = \tau^z \otimes \sigma^z, \quad 
  \label{eq:GammaModel}
\end{align}
so that the Hamiltonian becomes \cite{arovas_kitaev_shastry-sutherland}
\beq 
  \hlt_{\text{Kitaev}} = -\sum_{\gamma-\text{bonds}} J_\gamma \Gamma_j^\gamma \Gamma_k^\gamma,    \label{eq:kitaev}
\eeq 
where $\gamma = 1, \dots, 5$ labels the $j$--$k$ bond. Following Kitaev's original solution \cite{Kitaev2006anyons}, we represent the $\Gamma$-matrices in terms of  the aforementioned six Majorana operators by setting $\Gamma_j^\gamma = i a_j^\gamma c_j$. The Majoranas associated with the bonds can then be recombined into a $\intg_2$ gauge field $\hat{u}_{jk} \equiv i a_j^\gamma a_k^\gamma$ with eigenvalues ${u}_{jk} = \pm 1$. Like in the honeycomb Kitaev model, this $\intg_2$ gauge field is {\em static},
since all $\hat{u}_{jk}$ commute with the Hamiltonian. The relevant gauge-invariant quantities are the  $\intg_2$ fluxes through the elementary closed loops of the lattice (of length 4 and 3, respectively).
The first step in solving the model thus is to identify the ground-state configuration of these $\intg_2$ fluxes. 
Since the lattice at hand does not meet the requirements to apply Lieb's theorem~\cite{Lieb1994flux} to immediately 
identify the ground-state configuration, we instead  resort to a numerical exact solution of this problem via 
quantum Monte Carlo simulations, described in more detail below. 
The net result is that each loop of length 4 exhibits a $\pi$-flux, 
while for the triangular plaquettes (two of which add up to one 4-loop) the flux is  $\pm \pi/2$
\footnote{This flux assignment is also precisely what Lieb's theorem would dictate, were it applicable.}.
The two possible signs for the flux of the triangular plaquettes constitute time-reversed partners with equal energies -- one of the two has to be chosen in the resulting low-energy description of itinerant Majoranas coupled to a static $\intg_2$ gauge field
\footnote{This ground state of the $\intg_2$ fluxes is separated from all other flux configurations by a finite vison gap for all coupling parameters.}.

In general, the resulting free Majorana Hamiltonian has a particle-hole symmetry, which follows directly from the reality condition for Majorana fermions. While the original Hamiltonian also possesses a time-reversal symmetry, this is broken spontaneously by the ground state.
The system thus always resides in symmetry class D (instead of BDI for the honeycomb Kitaev model). With the systematic classification of TIs \cite{Schnyder2008classification,Kitaev2009periodic} in mind, symmetry class D allows for a $\intg$ invariant in two spatial dimensions, i.e. the occurrence of Chern insulators, as well as the possibility of a SOTI in the presence of a second-order lattice symmetry \cite{brouwer_HOTI_classification}.
Indeed, the Shastry-Sutherland lattice possesses two mirror symmetries along the diagonals of the rhombi (indicated by the dotted lines in Fig.~\ref{fig:panel}a), which are also symmetries of the Majorana Hamiltonian. In particular, the mirror operators $\mathcal{M}_{11}$ and $\mathcal{M}_{1\bar{1}}$ anticommute with the Hamiltonian as well as with each other (see also the Supplemental Material).

\begin{figure}[b]
    \centering
    \includegraphics[width=0.5\columnwidth]{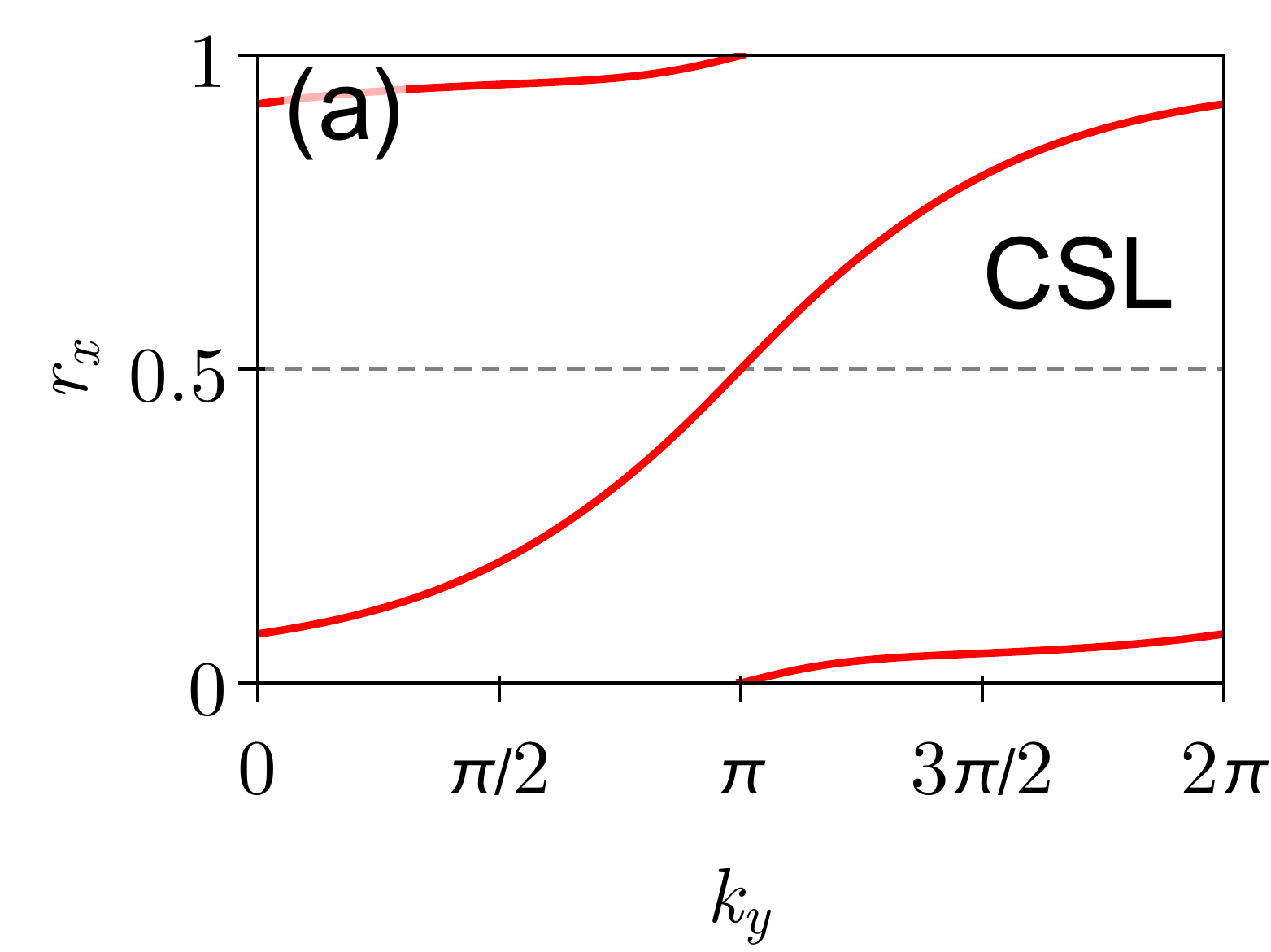}  \hspace{-0.02\textwidth}
    \includegraphics[width=0.5\columnwidth]{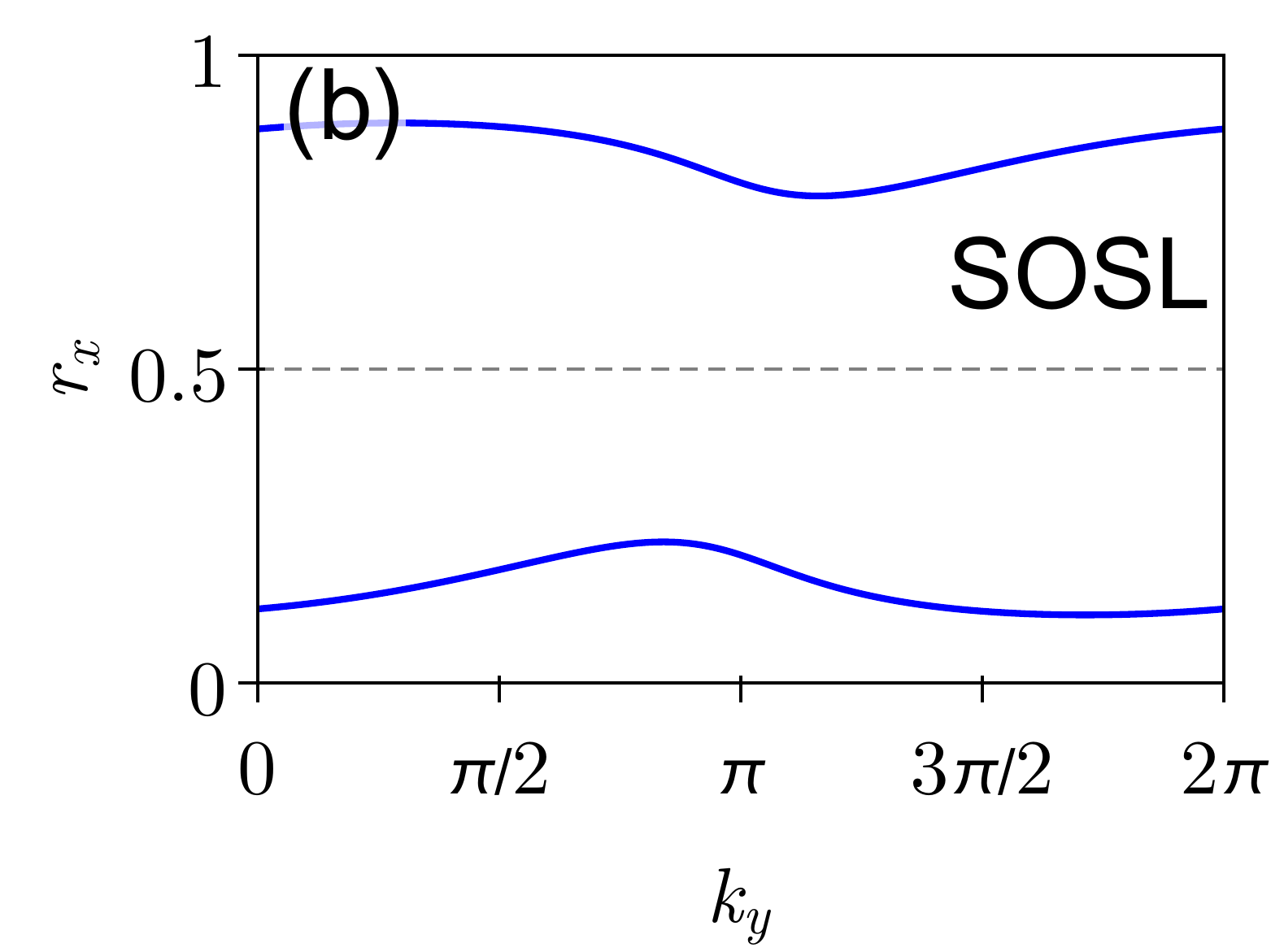}  \\ 
    \includegraphics[width=\columnwidth]{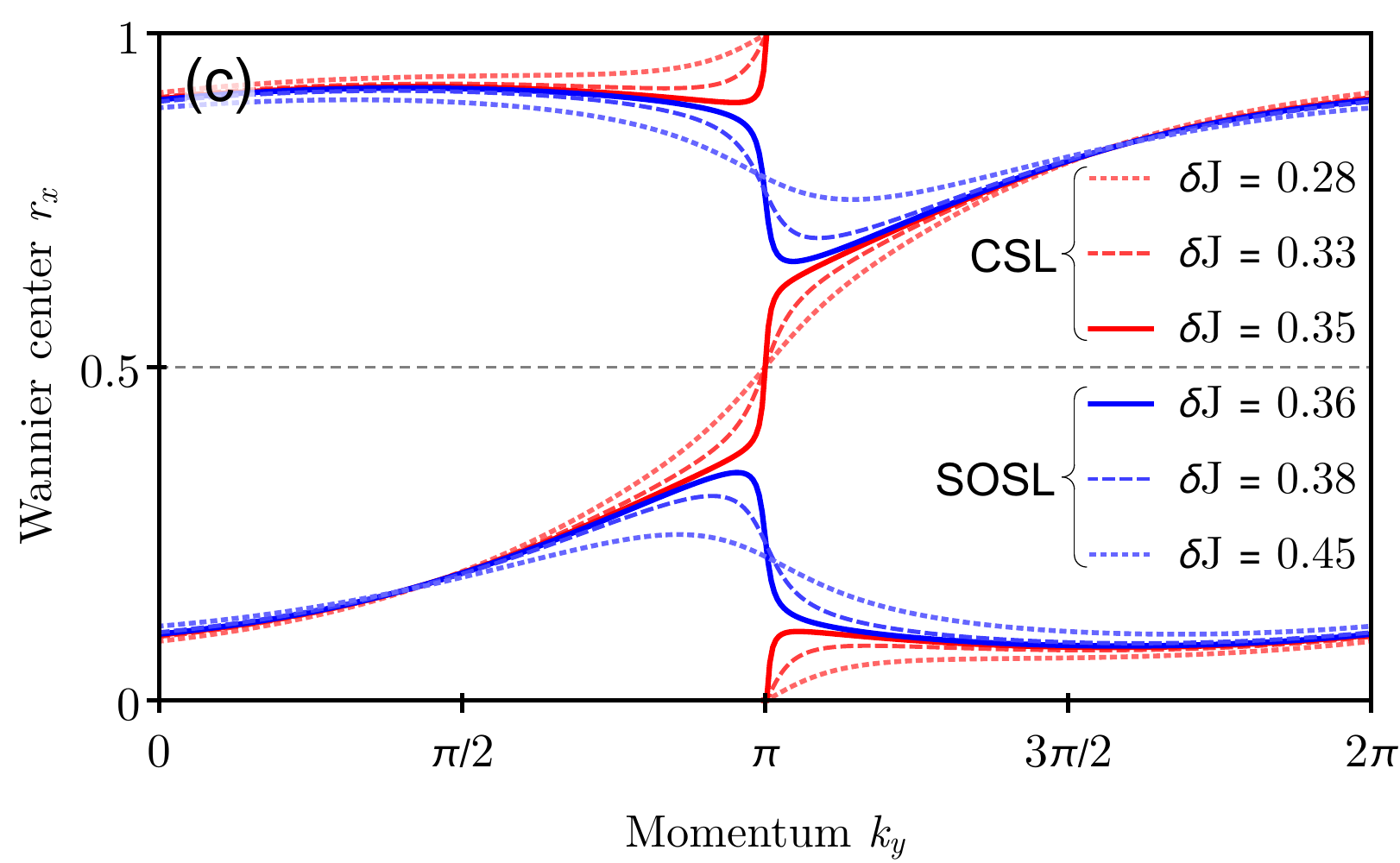}  
    \caption{
      {\bf Wannier bands} for  (a) the CSL ($\delta J = 0.2$) and (b) the SOSL ($\delta J = 0.5$) phase 
      with $J_0 = 0.8$, $J_z = 1$. 
      Since the Wannier centers 0 and 1 are equivalent, the plotted region is topologically a 2-torus. 
      The CSL is characterized by a winding of the Wannier bands along the torus, 
      while the SOSL is characterized by gapped Wannier bands along both $x$ and $y$ (not shown).
      (c) The transition between the CSL and SOSL phases. 
    }
    \label{fig:wannier}
\end{figure}

\paragraph{Ground-state phase diagram.--}
We can now proceed to discuss the ground-state phase diagram as a function of the coupling strength $J_0$ and the staggering $\delta J$, with $J_x = J_y = J_0 + \delta J$ and $J_x' = J_y' = J_0 - \delta J$, respectively. 
Following a Fourier transformation of the 4-band itinerant Majorana Hamiltonian, we obtain a bulk band structure, which is gapless  along the lines $J_z = \pm 2\sqrt{2} J_0$ [at $\vk = (0,0)$] and $J_z = \pm 2\sqrt{2} \delta J$ [at $\vk=(\pi,\pi)$], and gapped otherwise. The four resulting gapped phases are indicated in the phase diagram of Fig.~\ref{fig:panel}b).
We note that the phase diagram is reflection symmetric about the lines $J_0 = \pm \delta J$, since a reflection about these lines is equivalent to a $\intg_2$ gauge transformation. 

Computing the Chern number for the valence bands, we find a non-trivial Chern number of $+1$ for the valence band in two of these gapped phases, indicated by the red in the phase diagram. In terms of the Majorana fermions, these are conventional Chern insulators, while in the language of the original spin model, these phases constitute chiral spin liquids (CSLs). 
Discussed earlier \cite{arovas_kitaev_shastry-sutherland} in the context of the $\Gamma$-matrix model  \eqref{eq:kitaev}, these CSLs are higher-spin analogs of the CSL first discovered in a decorated honeycomb model by Yao and Kivelson \cite{yao-kivelson}. 
While the Chern number vanishes in the two remaining gapped phases, not both of them are trivial insulators. For sufficiently large staggering $\delta J$ (i.e. in the upper right corner of the phase diagram), we find a SOTI phase, which, in the language of the original spin model, can be referred to as a second-order spin liquid (SOSL). 
Computing the spectrum for the real space Hamiltonian on a square with open boundary conditions, we obtain four states near zero energy ($\ve = 0$), separated by a gap from the continuum. The corresponding wavefunctions are exponentially localized at the corners of the square, as shown in Fig.~\ref{fig:panel}c). We contrast this with the CSL, where we get a topologically protected chiral mode localized at the edge. We also observe that the SOSL  does not exhibit any zero modes on a system with periodic boundary conditions along one or both spatial directions.

The existence of corner modes is a hallmark of second-order TIs.
For the system at hand, these modes can be intuitively understood as a domain wall between two 1D topological phases \cite{brouwer_HOTI}. To wit, the system exhibits modes localized on mirror symmetric edges (i.e, along a diagonal in Fig.~\ref{fig:panel}a), which disperse along the edge and can be described by a $1+1$ dimensional massless Dirac Hamiltonian. 
For the (non mirror-symmetric) edges depicted in Fig.~\ref{fig:panel}a, these edge modes would gap out by addition of a mass term. 
However, since the two edges meeting at a corner are related by a mirror symmetry and the mass term must be odd under this symmetry, the corner is a mass domain wall in the Dirac Hamiltonian, which explains the presence of the corner mode. 

The presence of topologically protected corner modes can also be inferred from the \emph{bulk} bands by computing the Wannier centers \cite{hughes-bernevig_multipole_insulators}. 
More explicitly, we compute the hybrid Wannier functions \cite{vanderbilt_hybrid_wannier}, a basis of wavefunctions localized along the $x$ direction but delocalized along the $y$ direction (or vice versa), and plot the Wannier centers $r_x(k_y)$ modulo lattice translations \footnote{ 
  For details of this computation, see, for instance, Sec. II.A of Ref \cite{vanderbilt_wannier_sheets} or Sec. IV of Ref \cite{hughes-bernevig_multipole_insulators}. 
},  
so that $0$ and $1$ correspond to the same Wannier centers. 
The Wannier band topology can then be used to deduce the topological phase. 
For instance, for the CSL (CI), the Wannier band exhibits a nontrivial winding around the torus, while for the SOSL (SOTI), the Wannier bands are gapped along \emph{both} $x$ and $y$, with $r_{x,y}=1/2$ lying in the gap \cite{hughes-bernevig_multipole_insulators}. 
Since the mirror symmetries take $(x,y) \to \pm (y,x)$ (and anticommute with Hamiltonian), the Wannier centers along $y$ satisfy $r_y(k_x) = -r_x(k_y)$, so that $r_y$ is also gapped, thereby indicating a SOSL. 
These two distinct Wannier band topologies for the CSL and SOSL are plotted in Fig.~\ref{fig:wannier}a,b.
Finally, as we tune $\delta J$ to the CSL-SOSL transition, we clearly see a transition between these two scenarios, where a branch of the winding Wannier band of the CSL detaches and reattaches to a different branch to form the Wannier band structure of the SOSL, as shown in  Fig.~\ref{fig:wannier}c.

\paragraph{Thermodynamics.--}      

\begin{figure}[t]
    \centering
    \includegraphics[width=.95\columnwidth]{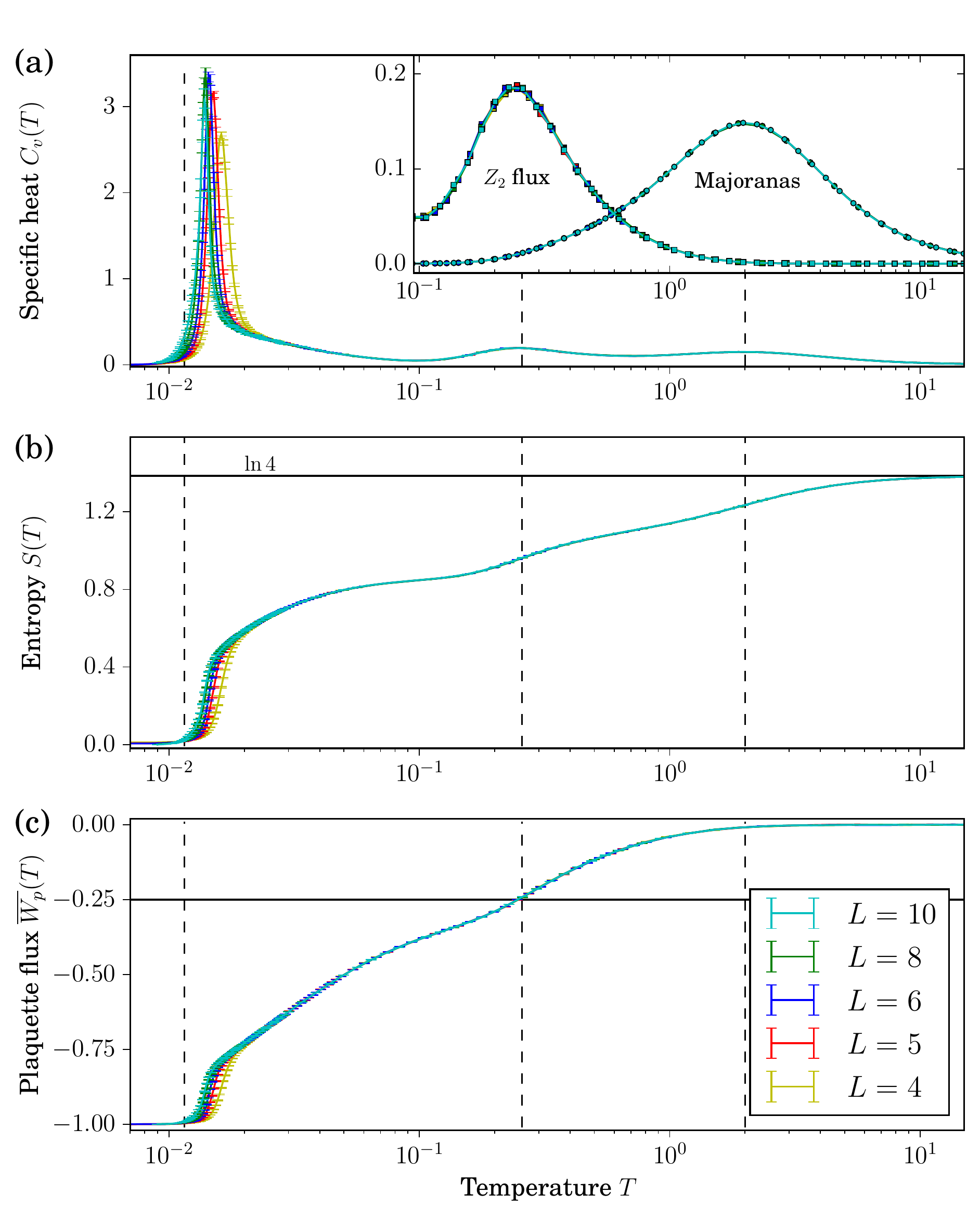}  
    \caption{
      {\bf Thermodynamics of the SOSL.}
      The upper panel shows the specific heat,
      the middle panel the entropy per spin, and 
      the lower panel the  $\intg_2$ flux per plaquette 
      as a function of temperature. 
      The three characteristic temperature scales for the low-temperature ordering transition 
      (extrapolated to the thermodynamic limit) and the two high-temperature crossovers
      are indicated by the dashed vertical lines.
    }
    \label{fig:qmc}
\end{figure}

To explore the thermal stability of the SOSL and the finite-temperature transition associated with spontaneous time-reversal symmetry breaking, we have employed large-scale quantum Monte Carlo simulations of our spin model, which are sign problem-free in the Majorana basis \cite{Nasu2014vaporization}. Our results, summarized in Fig.~\ref{fig:qmc}, indicate three relevant temperature scales, each one associated with a peak in the specific heat and a corresponding drop in the entropy per site. At the lowest temperature scale of $T\approx 0.01~J_z$ the system undergoes a phase transition, at which the $\intg_2$ gauge field orders into its ground-state configuration with a $\pi/2$-flux through all triangular plaquettes and a $\pi$-flux through all 4-loops, see Fig.~\ref{fig:qmc}c. A finite-size scaling analysis of the specific heat peak indeed reveals a divergence of the peak and a finite transition temperature $T_c = 0.012(1)~J_z$ in the thermodynamic limit (see Supplemental Material). Above this transition, we observe two independent thermodynamic crossovers, indicated by non-diverging peaks around $T_1 \approx 0.25~J_z$ and $T_2 \approx 2~J_z$ in the specific heat, see the inset of Fig.~\ref{fig:qmc}a. The higher crossover can be associated with the release of entropy of the Majorana fermions, whose energy scale is set by the hopping strength $J_z$, 
while the lower crossover is associated with a partial release of entropy of the $\intg_2$ gauge field due to the staggering $\delta J$.
The appearance of this lower crossover can be best understood by considering the limit $\delta J \rightarrow J_0$,
in which the system decomposes into decoupled 1D chains, formed by the dark gray plaquettes connected by the solid bonds in Fig.~\ref{fig:panel}a. At zero temperature, the individual chains are gapped and exhibit a $\pi$-flux per tetragonal plaquette \cite{Shankar_KitaevTC}. At finite temperature these fluxes, which constitute $1/4$ of the total flux of the 2D system, order at the temperature $T_1$, while the remaining plaquettes remain disordered. This results in a plateau in the $\intg_2$ flux per plaquette at $-1/4$, which clearly evolves as one approaches the 1D limit (as illustrated in Fig.~\ref{fig:supp2} of the Supplemental Material) and is accompanied by a smooth non-diverging peak in the specific heat (since this 1D physics does not give rise to a true phase transition). When $\delta J \neq J_0$ the remaining 3/4 of the plaquettes order at a much lower temperature scale, giving rise to the actual phase transition at $T\approx 0.01~J_z$.

It is interesting to note that, for the purely 1D tetragonal chain, the $\intg_2$ gauge field freezes into its ground state configuration at a temperature scale of order $\mathcal{O}(J)$. This is in marked contrast to the Kitaev model in two and three spatial dimensions where the same phenomenon occurs at $\mathcal{O}(10^{-2} J)$ \cite{Nasu2014vaporization,Nasu2015,Mishchenko2017}. The significantly higher temperature scale could have interesting experimental consequences for quasi-1D magnetic materials which realize Kitaev interactions. 

\paragraph{Discussion.--}    
The search for an experimental realization of the second-order Kitaev spin liquid and its clear thermodynamic signatures
of fractionalization at comparatively high temperature scales, could bring some diversity to the current hunt for Kitaev materials \cite{Trebst2017}. Here, a natural starting point is to first look for realizations of the Shastry-Sutherland lattice in spin-orbit dominated materials. One step in this direction has been taken by exploring the 4$f$ material DyB$_4$  \cite{Watanuki2005,Okuyama2005,Ji2007,Sim2016}, for which the spin-orbit coupling -- enhanced by the relatively high atomic number of $Z=66$ for Dy (compared to $Z=44/77$ for the Ru-/Ir-based Kitaev materials) -- holds promise to give rise to the required bond-directional exchange interactions. Given a suitable candidate material, the experimental detection of the corner modes of a second-order Kitaev spin liquid still poses a number of challenges. The density of states of the emergent Majorana fermions cannot be directly probed by scanning tunneling techniques, in contrast to conventional electronic systems \cite{Yazdani2014}. A more subtle experimental protocol is thus called for, perhaps taking advantage of the emergent quasiparticles' ability to carry heat. Indeed, the challenges mirror many of the problems of detecting emergent Majorana fermions in conventional Kitaev spin liquids, due to their lack of spin or charge quantum numbers. 

The study at hand complements previous theoretical work \cite{Obrien2016classification} on classifying topological band structures for {\em gapless} Majorana metals in two- and three-dimensional Kitaev models. Depending on the crystalline symmetries, these systems exhibit semimetals with Dirac \cite{Kitaev2006anyons} or Weyl points \cite{Hermanns2015weyl}, nodal lines \cite{Mandal2009exactly,Yamada2017} or topological metals with Majorana Fermi surfaces \cite{Hermanns2014quantum}.
Together with the present study this underpins the notion that Kitaev spin liquids can realize all known topological band structures in relatively simple and analytically tractable microscopic spin models. As such we expect that one can also construct Kitaev models that realize other higher-order spin liquids, including a SOSL with gapless hinges in three spatial dimensions, which we leave to future studies.

\vskip 5mm

\acknowledgments
We thank Jan Attig, Piet Brouwer, Max Geier, Victor Chua, Srinidhi Ramamurthy, Taylor Hughes and D. Khomskii for inspiration and useful discussions 
as well as Y. Motome and P. Mishchenko for collaboration on a related numerical project. 
We acknowledge partial support from the Deutsche Forschungsgemeinschaft (DFG) within the CRC network TR 183 (projects B01 and B03)
and SFB 1238 (project C03).

\bibliography{ss_lattice}

\pagebreak
\widetext

\begin{center}
  \textbf{\large Supplemental Material}
\end{center}

\setcounter{section}{0}
\setcounter{equation}{0}
\setcounter{figure}{0}
\setcounter{table}{0}
\setcounter{page}{1}
\makeatletter
\renewcommand{\theequation}{S\arabic{equation}}
\renewcommand{\thefigure}{S\arabic{figure}}

\section{Lattice model}
The Shastry-Sutherland lattice can be constructed from a 2D square latticeby adding diagonal bonds in every other square. A more ``symmetric'' version of this lattice can be constructed by deforming the squares into rhombuses with corner angle $\theta$. The lattice has a 4 site unit cell, with lattice positions 
\begin{align} 
  \vr_1 = (0, \, 0), \quad \vr_2 = \frac{1}{2} (1, -a), \quad \vr_3 = \frac{1}{2}(b, \, b),  \quad \vr_4 = \frac{1}{2} (a, -1),
\end{align}
where $a = \tan \left( \frac{\pi}{4} - \frac{\theta}{2} \right)$ and $b = \sqrt{2} \cos \left( \frac{\theta}{2} \right) \, \sec \left( \frac{\pi}{4} - \frac{\theta}{2} \right)$. 
The square lattice is recovered for $\theta = \pi/2$. The lattice possesses two mirror symmetries along the diagonals, a twofold rotation symmetry about the diagonal bond center, a fourfold rotation symmetry about the centers of the empty squares, as well as two glide symmetries.

The generalized Kitaev model of eq \eqref{eq:kitaev} is solved by decomposing $\Gamma$'s into six Majoranas, as $\Gamma_j^\gamma = i a_j^\gamma c_j$. This doubles the dimension of the Hilbert space, and the physical Hilbert space is the eigenvalue $+1$ sector of the operator $D_j = ia_j^1 a_j^2 a_j^3 a_j^4 a_j^5 c_j$ for each site $j$. The ground state flux configuration of the $\intg_2$ gauge field, \viz, $\pi$-flux through the 4-loops and $\pi/2$ through the 3-loops, is realized by setting $u_{jk} = \langle i a_j^\gamma a_k^\gamma \rangle = 1$ whenever $j$ is a lower-numbered site than $k$. The resulting itinerant Majorana Hamiltonian is 
\begin{align}
 \hlt = & \; i\sum_{m,n} \left[ J_x \left( c_{m,n,1} c_{m-1,n,2} + c_{m,n,3} c_{m-1,n,4} \right) + J_y \left( c_{m,n,1} c_{m,n-1,4} + c_{m,n,2} c_{m-1,n,3}  \right)   \right. \nonumber \\ 
 & \; + J_x' \left( c_{m,n,1} c_{m,n,2} + c_{m,n,3} c_{m,n,4} \right) + J_y' \left( c_{m,n,1} c_{m,n,4} + c_{m,n,2} c_{m,n,3}  \right) - \left. J_z \left( c_{m,n,2} c_{m,n,4} + c_{m,n,1} c_{m+1,n-1,3} \right) \right] \nonumber 
\end{align}
By a Fourier transform, we get a Bloch Hamiltonian for a 4-band model:
\beq 
  \hlt = i 
  \begin{pmatrix}
      0				& J_x' + J_x \e^{-i k_x}	& -J_z \e^{-i(k_x-k_y)}		&  J_y' + J_y \e^{ik_y}		\\
      -J_x' - J_x \e^{ik_x}	& 0				& J_y' + J_y \e^{ik_y}		&  -J_z				\\
      J_z \e^{i(k_x-k_y)}	& -J_y' - J_y \e^{-i k_y}	& 0				&  J_x' + J_x \e^{i k_x}	\\
      -J_y' - J_y \e^{-i k_y}	& J_z				& -J_x' - J_x \e^{-i k_x}	&  0		
  \end{pmatrix},
\eeq
where $J_x = J_y = J_0 + \delta J$ and $J_x' = J_y' = J_0 - \delta J$. The unitary operators for the mirror symmetries along the $11$ and $1\bar{1}$ directions are
\beq 
  \mathcal{M}_{11} = 
    \begin{pmatrix}
      0  &  0 &  1 &  0 \\
      0  &  1 &  0 &  0 \\
      1  &  0 &  0 &  0 \\
      0  &  0 &  0 & -1 
    \end{pmatrix}, \qquad
  \mathcal{M}_{1\bar{1}} = 
    \begin{pmatrix}
      -1 &  0 &  0 &  0 \\
      0  &  0 &  0 &  1 \\
      0  &  0 &  1 &  0 \\
      0  &  1 &  0 &  0 
    \end{pmatrix}.
\eeq
It can be explicitly checked that these operators satisfy the anticommutation relation $\, \{\mathcal{M}_{11}, \mathcal{M}_{1\bar{1}}\} = 0$.

\section{Numerical analysis}

\paragraph{Monte Carlo approach.--}
For our numerical analysis of thermodynamic observables, we have employed large-scale quantum Monte Carlo simulations,
which in the Majorana basis are sign-problem free. In this approach, which has been spearheaded in Ref.~\onlinecite{Nasu2014vaporization}, one samples configurations $\{u_{jk}\}$ of the $\mathbb{Z}_2$ gauge field with the statistical weight for each configuration calculated via an exact diagonalization of the Majorana fermions. 
Specifically, the Hamiltonian in a fixed gauge field configuration is diagonalized to a canonical form \cite{Kitaev2006anyons}
$
\mathcal{H} = \sum_{\lambda = 1}^{N/2} \epsilon_\lambda \left(a_\lambda^\dagger a_\lambda^{\phantom\dagger} - \frac{1}{2} \right),
$
where $N$ denotes the number of spins in the system, while $a_\lambda^\dagger$, $a_\lambda^{\phantom\dagger}$ are the creation and annihilation operators of spinless fermions, each one composed of two itinerant Majorana modes.  
The partition function of the full system can be written as
\begin{equation}
\mathcal{Z} = \tr_{\{u_{jk}\}} \tr_{\{c_i\}} e^{-\beta \mathcal{H}} = \tr_{\{u_{jk}\}} e^{-\beta F(\{u_{jk}\})} \,,
\end{equation}
where $F(\{u_{jk}\})$ denotes the free energy of the itinerant Majorana fermions in a given $\mathbb{Z}_2$ gauge field configuration. The free energy $F$ and all other thermodynamic observables are derived from the partition function of the Majorana system in a fixed $\{u_{jk}\}$ which is obtained via the explicit summation over all fermionic Fock states
$
\mathcal{Z}_{\{c_i\}} = \prod_{\lambda = 1}^{N/2} 2 \cosh \left( \frac{\beta \epsilon_\lambda}{2} \right ).
$
The energy of the Majorana system is then given by
$
E_F (\{u_{jk}\}) = - \sum_\lambda \frac{\epsilon_\lambda}{2} \tanh \left( \frac{\beta \epsilon_\lambda}{2} \right) \,.
$
In order to separate the specific heat contribution of the itinerant Majorana fermions from the fluctuations in the $\mathbb{Z}_2$ gauge field, we calculate \cite{doi:10.1143/JPSJ.74.661}
\begin{align}
C_{v, F} (T) &= \frac{1}{T^2} \left( \langle E^2 (\{u_{jk}\}) \rangle_F - \langle E(\{u_{jk}\}) \rangle_F^2 \right) 
= \frac{1}{T^2} \sum_\lambda \frac{\epsilon_\lambda^2}{4} \left( 1 - \tanh^2 \left(\frac{\beta \epsilon_\lambda}{2} \right) \right) 
 = - \frac{1}{T^2} \frac{\partial E_f(\{u_{jk}\})}{\partial \beta} \,,
\end{align}
which gives a total specific heat of
\begin{equation}
C_v(T) = \frac{1}{T^2} \left( \underbrace{\left \langle E_F^2(\{u_{jk} \}) \right \rangle_{MC} - \left \langle E_F(\{u_{jk} \}) \right \rangle_{MC}^2}_{\text{Gauge field contribution}} - \underbrace{ \left \langle \frac{\partial E_f(\{u_{jk}\})}{\partial \beta} \right \rangle_{MC} }_{\text{It. Majorana contribution}}\right) \,.
\end{equation}

Finally, we note that in deriving the Majorana partition function, we have not distinguished between physical and unphysical fermionic Fock states. It is well known that a given $\mathbb{Z}_2$ gauge field configuration on a system with certain boundary conditions allows for either even or odd fermionic parity states \cite{Pedrocchi2011}, with only one of the two constituting the physical states. The unphysical states, which correspond to states of the expanded Hilbert space, contribute deviations of order $1/N$ \cite{Zschocke2015} to observables and can be neglected in the thermodynamic limit.

\paragraph{Simulation setup.--}
All the simulations were performed on systems with periodic boundary conditions. To avoid the slowing down and freezing of the Monte Carlo sampling at low temperatures, we employed parallel tempering with 24 - 64 replicas in each simulation. 
For all the systems, 20,000 measurement sweeps were performed (after 10,000 thermalization sweeps), with every sweep being followed by an attempted replica exchange.  

\begin{figure*}[h]
   \centering
    \includegraphics[ width=0.42\textwidth]{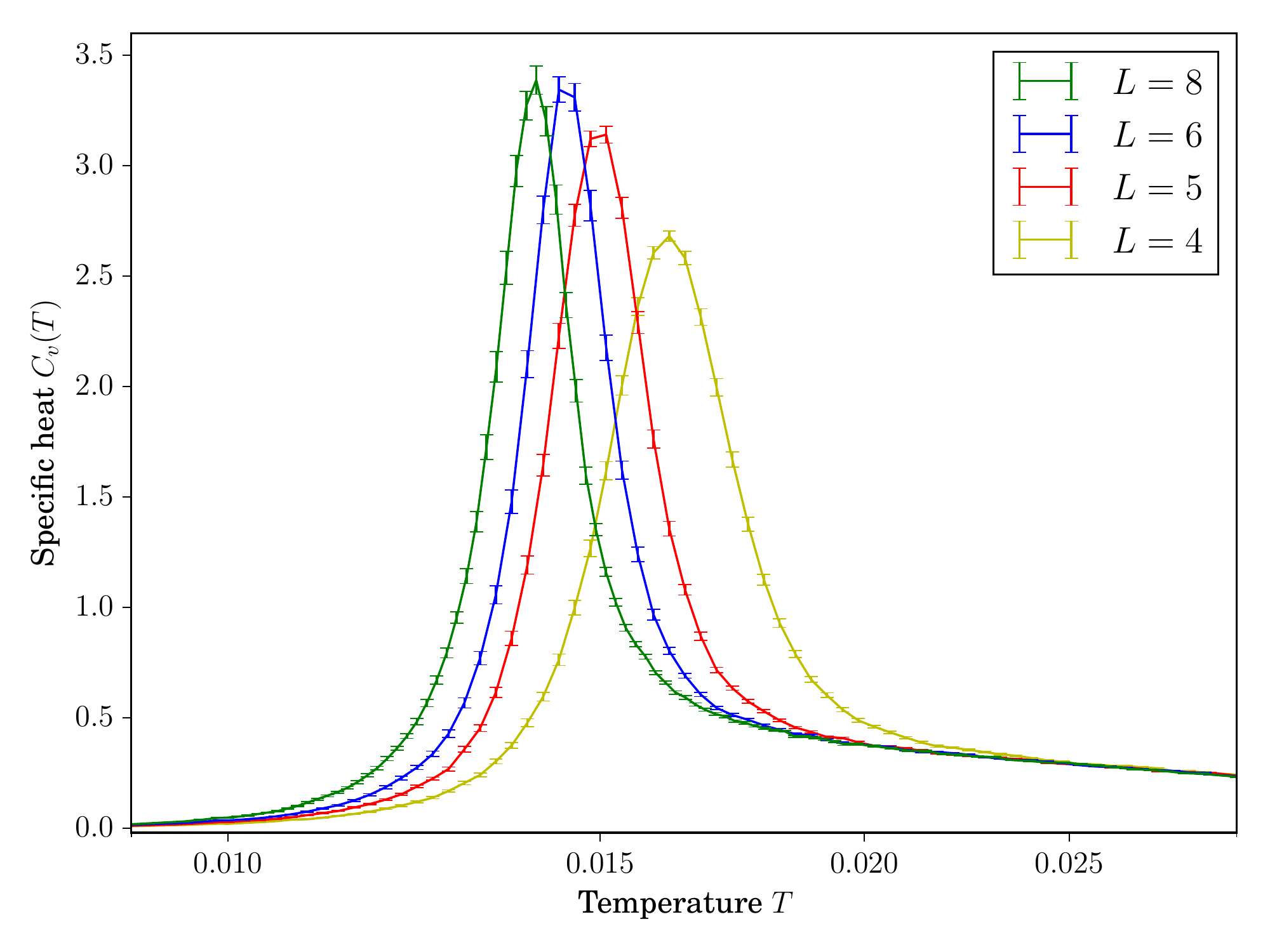}
    \hspace{0.08 \textwidth}
    \includegraphics[ width=0.42\textwidth]{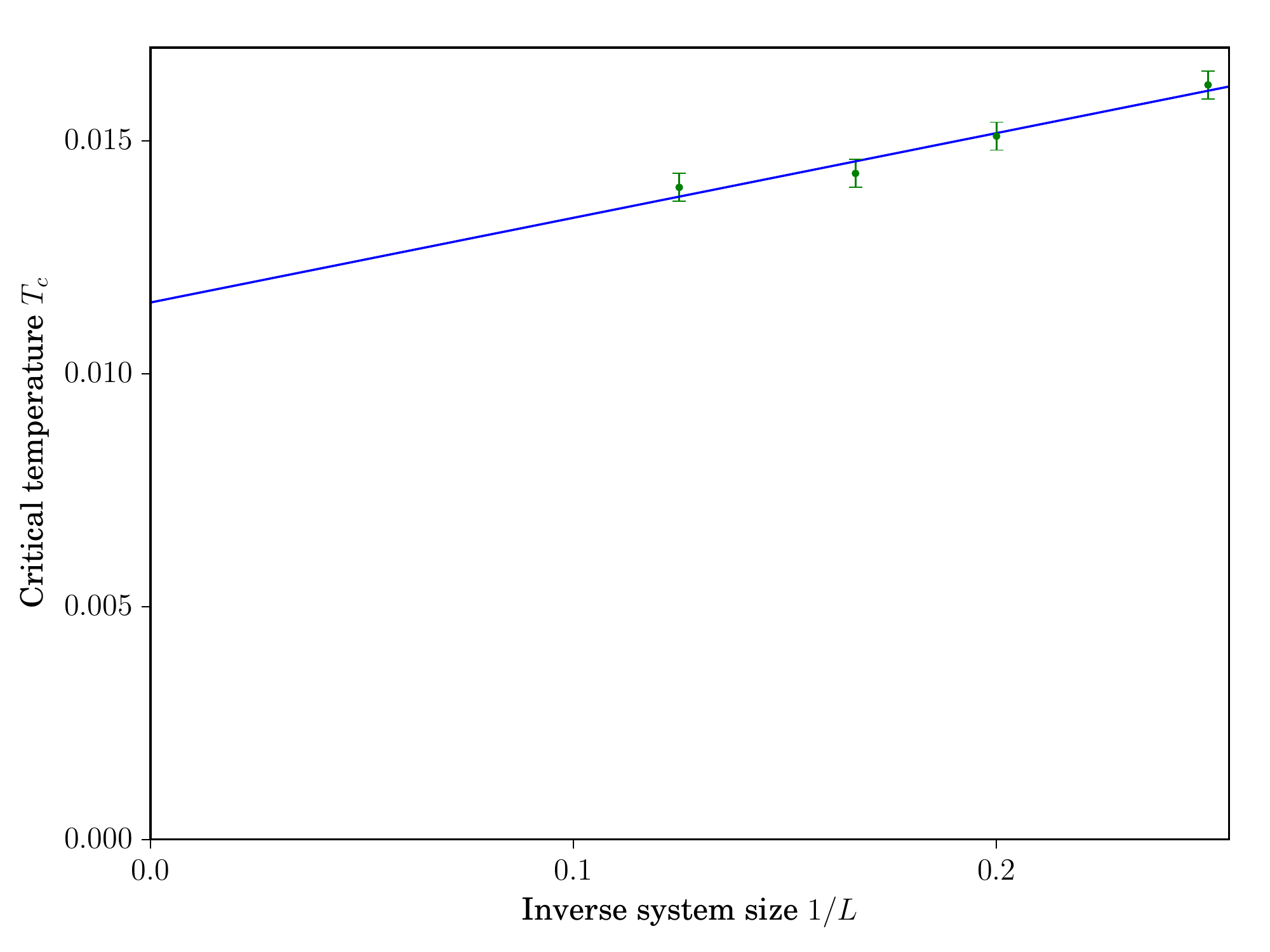}
    \caption{
      {\bf Finite-size scaling analysis.}
      (a) Gauge field contribution to the specific heat around the low-temperature peak for different system sizes $L$ 
      (b) Scaling plot of the peak position versus the inverse system size $1/L$. The solid line indicates a linear fit.} 
    \label{fig:supp1}
\end{figure*}

\paragraph{Results.--}
The estimate for the critical temperature in the thermodynamic limit was obtained from linear extrapolation of the position of the low-temperature peak of the specific heat versus the inverse system size $1/L$, as illustrated in Fig.~\ref{fig:supp1}~b).

\begin{figure*}[h]
   \centering
   \includegraphics[ width=0.42\textwidth]{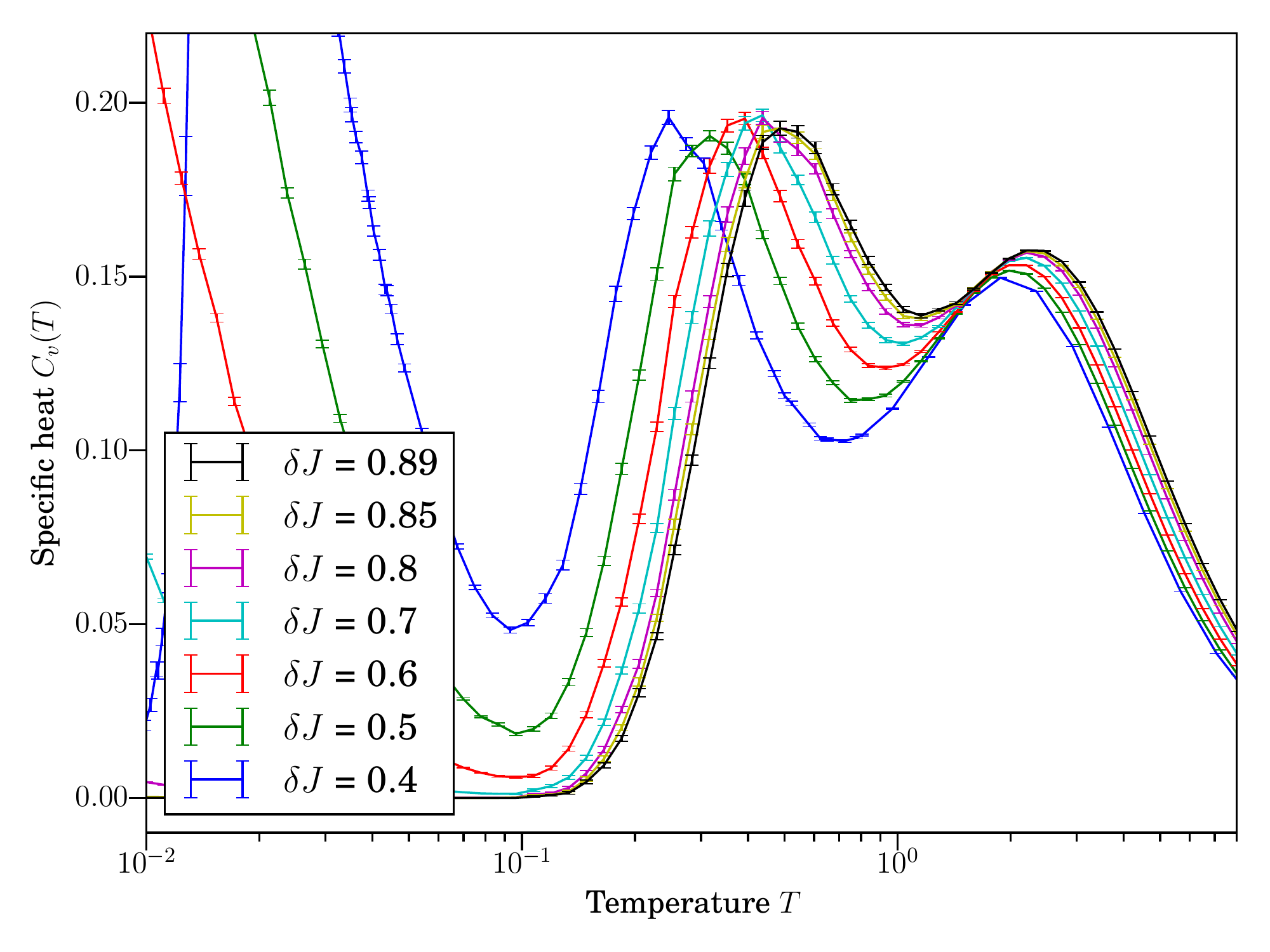}
    \hspace{0.08 \textwidth}
   \includegraphics[ width=0.42\textwidth]{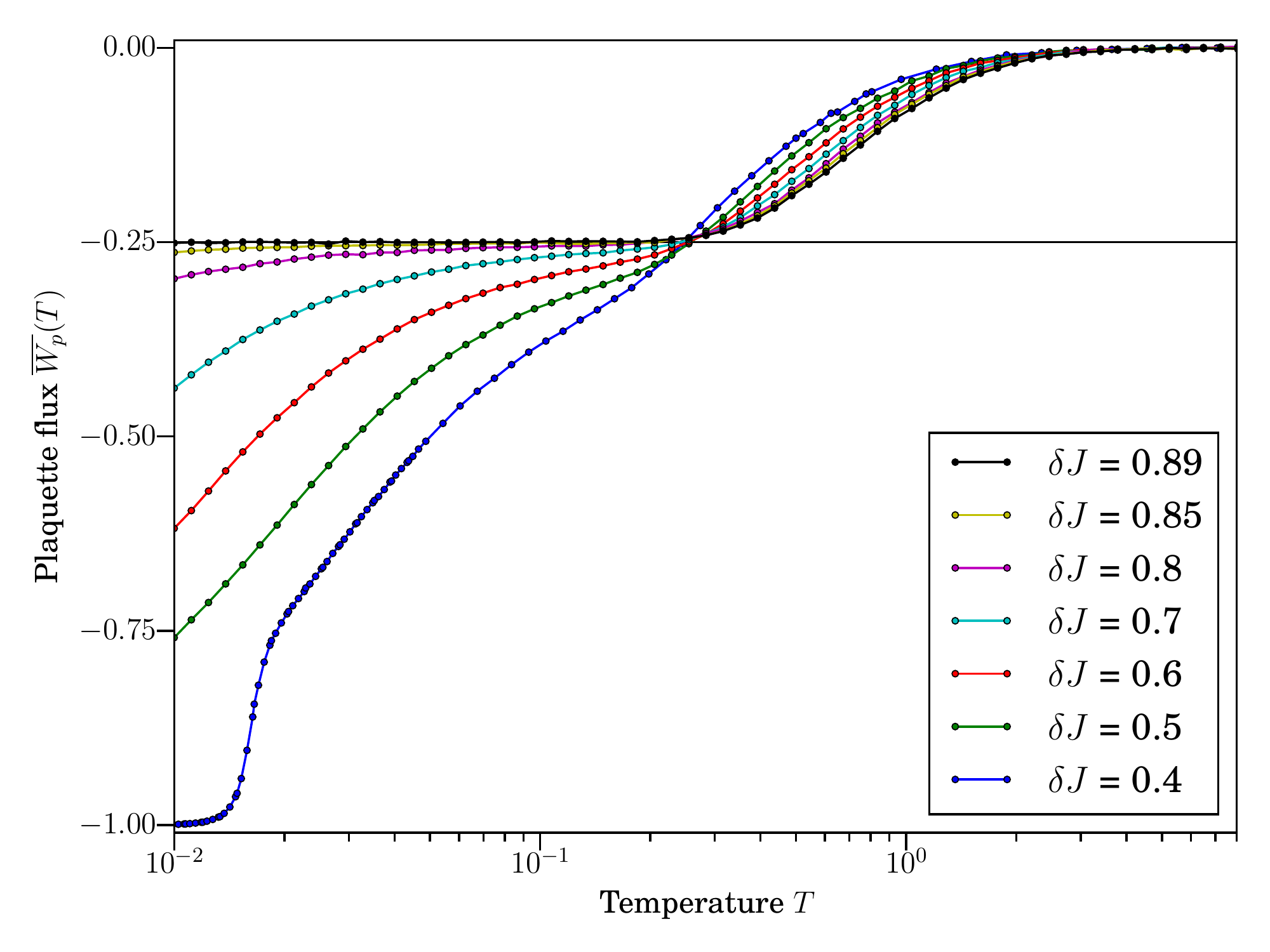}
    \caption{
      {\bf Approaching the 1D limit.}
      (a) Specific heat and (b) plaquette flux as one approaches the 1D limit $\delta J \to J$ for $J_0 = 0.9$.
	}
    \label{fig:supp2}
\end{figure*}

The approach to the 1D limit of the model for $\delta J \to J$ is illustrated in Fig.~\ref{fig:supp2}, which shows 
the specific heat and plaquette flux for $J_0 = 0.9$ and different values of $\delta J$. 
While the low-temperature crossover of the specific heat peak wanders towards $T_1 \approx 0.55 J_z$,
a plateau at $-1/4$ forms in the plaquette flux (as described in the main text).

\end{document}